# Anomaly Detection in Streaming Sensor Data

Alec Pawling
University of Notre Dame, USA

Ping Yan
University of Notre Dame, USA

Julián Candia
Northeastern University, USA

Tim Schoenharl
University of Notre Dame, USA

Greg Madey
University of Notre Dame, USA

## Abstract

In this chapter we consider a cell phone network as a set of automatically deployed sensors that records movement and interaction patterns of the population. We discuss methods for detecting anomalies in the streaming data produced by the cell phone network. We motivate this discussion by describing the Wireless Phone Based Emergency Response (WIPER) system, a proof-of-concept decision support system for emergency response managers. We also discuss some of the scientific work enabled by this type of sensor data and the related privacy issues. We describe scientific studies that use the cell phone data set and steps we have taken to ensure the security of the data. We describe the overall decision support system and discuss three methods of anomaly detection that we have applied to the data.

## Keywords

## Introduction

The Wireless Phone-Based Emergency Response System (WIPER) is a laboratory proof-of-concept, Dynamic Data Driven Application System (DDDAS) prototype that uses cell phone network data to identify potential emergency situations and monitor aggregated population movement and calling activity. The system is designed to complement existing emergency response management tools by providing a high level view of human activity during a crisis situation using real-time data from the cell phone network in conjunction with geographical information systems (GIS). Using cell phones as sensors has the advantages of automatic deployment and sensor maintenance; however, the data available from the network is limited. Currently only service usage data and coarse location data, approximated by a Voronoi lattice defined by the cell towers, are available, although cell-tower triangulation and GPS could greatly improve the location data (Madey, Szabó, & Barabási, 2006, Madey et al., 2007, Pawling, Schoenharl, Yan, & Madey, 2008, Schoenharl, Bravo, & Madey, 2006, Schoenharl, Madey, Szabó, & Barabási, 2006, Schoenharl & Madey, 2008).

The viability of using cell phones as a sensor network has been established through the use of phone location data for traffic management (Associated Press, 2005). WIPER applies this finding to fill a need in emergency response management for a high level view of an emergency situation that is updated in near real-time. Tatomir and Rothkrantz (2005) and Thomas, Andoh-Baidoo, and George (2005) describe systems for gathering on-site information about emergency situations directly from response worker on the ground via ad-hoc networks of PDAs. While these systems can provide detailed information about some aspects of the situation, such as the location of victims and environmental conditions, the information is limited to what can be observed and reported by the responders. This provides valuable but local information, though there may be observations from different, geographically dispersed locations. In contrast, WIPER provides less detail, but instead gives an overall view of population movements that may be valuable in refining the response plans or directing response workers to gather more detailed information at a particular location.

Dynamic data driven applications systems (DDDAS) provide a framework in which running simulations incorporate data from a sensor network to improve accuracy. To achieve this, the simulations dynamically steer the measurement process to obtain the most useful data. The development of DDDAS applications is motivated by the limited ability to predict phenomena such as weather and wildfire via simulation. Such phenomena are quite complex, and the correct simulation parameterization is extremely difficult. The goal of DDDAS is to provide robustness to such simulations by allowing them to combine sub-optimal initial parameterizations with newly available, real-world data to improve performance without the expense of rerunning the simulations from the beginning (Douglas & Deshmukh, 2000).

In this chapter, we focus on one component of WIPER: the detection and alert system. This module monitors streaming data from the cell phone network for anomalous activity. Detected anomalies are used to initiate an ensemble of predictive simulations with the goal of aiding emergency response managers in taking effective steps to mitigate crisis events. We discuss methods for anomaly detection on two aspects of the call data: the call activity (the number of calls made in a fixed time interval) and the spatial distribution of network usage.

The remainder of the chapter is organized as follows: we discuss background literature related to mining data from a cell phone network. We start with a discussion of methods for detecting outliers in our data, with a focus on using data clustering to model normality in data.

Those clusters of outliers in the streaming data could be indicators of a problem, disaster or emergency in a geographical area (e.g., an industrial explosion, a civil disturbance, progress of a mandated evacuation prior to a hurricane, a terrorist bombing)  We then give an overview of the data set and the WIPER system, followed by descriptions of algorithms used in the detection and alert system. Finally, we discuss some of the privacy issues related to this work and our plans for future work in the spatial, graph and temporal analysis of the underlying social network.

## Background

In this section, we discuss background literature on outlier detection and clustering especially that relevant to our application of detecting anomalies in streaming cell phone sensor data: both 1) location and movement data and 2) calling patterns of the population carrying the cell phones.

### *Outlier Detection*

An outlier is an item in a data set that does not appear to be consistent with the rest of the set (Barnett & Lewis, 1994). There is a great deal of literature on the problem of outlier detection as well as a number of applications, including fraud detection, intrusion detection, and time series monitoring (Hodge & Austin, 2004).

There are three fundamental approaches for outlier detection:

- *Model both normality and abnormality*: this approach assumes that a training set representative of both normal and abnormal data exists.
- *Model either normality or abnormality*: this approach typically models normality and is well suited for dynamic data.
- *Assume no a priori knowledge of the data*: this approach is well suited for static

distributions and assumes that outliers are, in some sense, far from normal data. (Hodge & Austin, 2004).

Additionally, there are four statistical techniques for outlier detection: parametric, semi-parametric, non-parametric, and proximity based methods. Parametric outlier detection techniques assume that the data follows a particular probability distribution. These techniques tend to be fast but inflexible. They depend on a correct assumption of the underlying data distribution and are not suitable for dynamic data. Semi-parametric models use mixture models or kernel density estimators rather than a single global model. Both mixture models and kernel models estimate a probability distribution as the combination of multiple probability distributions. Non-parametric techniques make no assumptions about the underlying distribution of the data and tend to be computationally expensive. Proximity based techniques define outliers in terms of their distance from other points in the data set and, like non-parametric techniques, make no assumptions about the data distribution (Hodge & Austin, 2004).

In this chapter, we approach the outlier detection problem by modeling normal behavior. One technique for modeling normality in multidimensional space is data clustering, which enables outlier detection using proximity based techniques.

## *Data Clustering*

The goal of data clustering is to group similar data items together. Often, similarity is defined in terms of distance: the distance between similar items is small. Data items that do not belong to any cluster, or data items that belong to very small clusters, may be viewed as outliers, depending on the clustering algorithm and application.

The clustering problem is defined as follows: let a data set $D$ consist of a set of data items

$\{\vec{d}_1, \vec{d}_2, \ldots\}$ such that each data item is a vector of measurements, $\vec{d}_i = \langle d_{i,1}, d_{i,2}, \ldots, d_{i,n} \rangle$. Clustering provides a convenient way for finding anomalous data items: anomalies are the data items that are far from all other data items. These may be data items that belong to no cluster, or they may be the data items that belong to small clusters. If we take the view that anomalies belong to no cluster, we can use a clustering of the data to model normal behavior. If we view each cluster as a single point, we can greatly reduce the cost of proximity based anomaly detection, assuming the number of clusters is small relative to the total number of data items and that we can cluster the data quickly.

Traditional clustering algorithms can be divided into two types: partitional and hierarchical. Partitional algorithms, such as *k*-means and expectation maximization, divide the data into some number, often a predefined number, of disjoint subsets. These algorithms often start with a random set of clusters and iterate until some stopping condition is met. As a result, these algorithms have a tendency to converge on local minima. Hierarchical algorithms divide the data into a nested set of partitions and are useful for discovering taxonomies in data. They may either take a top-down approach, in which an initial data cluster containing all of the data items is iteratively split until each data item is in its own cluster, or a bottom-up approach, in which clusters initially consisting of only a single element are iteratively merged until all of the data items belong to a single cluster. Often, hierarchical algorithms must compute the distance between each pair of data items in the data set, and, therefore, tend to be computationally expensive, though there are techniques for making this process more efficient (Jain, Murty, & Flynn, 1999).

Partitional and hierarchical clustering algorithms may also incrementally incorporate new data into the cluster model (Jain et al, 1999). The leader algorithm incrementally partitions the data into cluster using a distance threshold to determine if a new data item should be added to an existing cluster or placed in a new cluster (Hartigan, 1975). Fisher (1987) describes the COBWEB algorithm, an incremental clustering algorithm that identifies a conceptual hierarchy. The algorithm uses category utility, a function that provides a measure of similarity of items in the same cluster and dissimilarity of items in different cluster, to determine whether a new object should be classified using an existing concept in the hierarchy or whether a new concept should be added. The COBWEB algorithm also combines and splits classes as necessary, based on category utility. Charikar et al (1997) describe several incremental agglomerative clustering algorithms for information retrieval applications. In these algorithms, when a new data item does not meet the criteria for inclusion in one of the existing clusters, a new cluster is created and two other clusters are merged so that *k* clusters exist at all times. The algorithms differ in their approach to determining the two clusters to be merged.

Stream clustering algorithms are similar to incremental algorithms. In addition to processing each item only once, stream algorithms typically use no more that order polylogarithmic memory with respect to the number of data items. Guha, Meyerson, Mishra, Motwani, and O'Callaghan (2003) present a method based on *k*-medoids—an algorithm similar to *k*-means. The clusters are computed periodically as the stream arrives, using a combination of the streaming data and cluster centers from previous iterations to keep memory usage low. Aggarwal, Han, Wang, and Yu (2003) present a method that takes into account the evolution of streaming data, giving more importance to more recent data items rather than letting the clustering results be dominated by a significant amount of outdated data. The algorithm computes *micro-clusters*, which are statistical summaries of the data, periodically throughout the stream. These micro-clusters serve as the data points for a modified *k*-means clustering algorithm.

Hybrid techniques combine two clustering algorithms. Cheu, Keongg, and Zhou (2004)

examine the use of iterated, partitional algorithms, such as *k*-means, as a method of reducing a data set before applying hierarchical algorithms. Chipman and Tibshirani (2006) combine agglomerative algorithms, which tend to effectively discover small clusters, with divisive methods, which tend to effectively discover large clusters. Surdeanu, Turmo, and Ageno (2005) propose a hybrid clustering algorithm that uses hierarchical clustering to determine initial parameters for expectation maximization.

### *Percolation Theory*

We use concepts from percolation theory to discover spatial anomalies. Percolation theory studies the emergence of connected components, or clusters, in a *d*-dimensional lattice as the probability of an edge existing between a pair of neighbors in the lattice approaches 1. At some critical probability a percolating cluster, a connected component containing most of the vertices in the lattice, appears. Percolation theory is typically interested in three quantities: the fraction of the lattice in the percolating cluster, the average cluster size, and the cluster size distribution (Stauffer & Aharony, 1992, Albert & Barabási, 2002).

## The Dataset

We use a large primary dataset consisting of actual call record data from a cellular service provider to support the development of WIPER. The data set contains records describing phone calls and SMS messages that traverse the network. Each record contains the date, time, the ID of the individual making the call, the ID of the individual receiving the call, the tower the caller's phone is communicating with, and the type of transaction: voice call or SMS. The IDs of the individuals making and receiving the calls are generated by the service provider using an undisclosed hash function. Currently, we have approximately two years of call record data, taking up 1.25 TB of disk space after being compressed with gzip. One month of data contains approximately five hundred million records. Roughly one quarter of these are SMS transactions, and the remaining three-quarters are voice phone calls. An additional dataset provides the latitude and longitude of the cell towers. In addition to the primary dataset (actual call record data) from the service provider, we generate a secondary data set (synthetic call record data) using validated simulations from a component of the WIPER system. Although there are many anomalies in the primary dataset of actual call data, in most cases we have not been able to determine the cause (although highly newsworthy events have been correlated with anomalies in the data streams). The synthetic data helps us test our anomaly detection algorithms and other components of the WIPER system in a more controlled manner.

Since the call record includes the location and the time of the transactions (voice and SMS), when aggregated, it forms a times series data stream with normal patterns varying with day of the week, time of the day, and cell-tower location. Abnormal patterns in the data, what we call anomalies, could be indications of a disaster, e.g., an airplane crash, a political riot, or a terrorist attack. Of course the anomaly could be caused by a more benign event such as traffic after a championship football game, a public holiday, or a news event outside the area under watch. Such anomalous patterns could reflect many individuals making phone calls or sending SMS text messages because of a traffic jam, a public holiday, or a nearby disaster. In all cases the level of calling would increase above a baseline and be visible in the time series data streams as an anomaly.

The primary data has been used in a number of other studies. Onnela et al. (2007a) analyze a wide range of characteristics of the call graph, including degree distribution, tie strength distribution, topological and weighted assortativity, clustering coefficient, and percolation properties.

In a second study, Onnela et al. (2007b) explore the role of tie strength in information diffusion. A graph is built from 18 months of data, using the total call duration between pairs of users as the edge weights. The analysis of the graph shows a positive correlation between the strength of an edge and the number of common neighbors shared by the two vertices, indicating that strong ties tend to form within community structures and that weak ties form the connections between communities. To measure the importance of strong and weak ties, two experiments are performed: the edges are removed in increasing order of strength and the edges are removed in decreasing order of strength. Removing weak ties causes the graph to collapse into many small components very quickly, where removing the strong ties causes the graph to shrink slowly.

The usefulness of this data goes beyond social network analysis and the development of emergency response tools. González, Hidalgo, & Barabási (2008) study human movement patterns over a six month period. Information from this type of study can be used for a number of applications, including design of public transportation systems, traffic engineering, and prediction and control of disease outbreak. Candia et al (2008) and González & Barabási (2007) discuss the privacy implications of working with this type of data in the context of scientific research.

## WIPER – Cell Phones as Sensors for Situational Awareness

One goal of the WIPER project is to develop a laboratory proof-of-concept to evaluate the potential of using cell phones as sensors to increase situational awareness of emergency response managers during an ongoing crisis. The WIPER system is designed to accept streams of near real-time aggregated data about calling activity and location data of the cell phones in a geographical area. This data could be monitored for anomalies that could serve as alerts of potential problems or emergencies (the primary focus of this chapter), but could also be displayed on maps to provide emergency mangers a view of where the citizens are, their movements, and potential “hot spots” indicated by above normal calling activity. The system runs simulations that attempt to infer the nature of the anomalous event and to predict future behavior of the cell phone network and, hence, the population affected by the crisis. New data is used as it becomes available from the cell phone network to validate and steer running simulations in order to improve their predictive utility.

The WIPER system consists of five components, each of which is described briefly below.

- The *Decision Support System* (DSS) is a web-based front end through which emergency response managers interact with the WIPER system.
- The *Detection and Alert System* (DAS) monitors streaming network data for anomalous activity. There are various aspects of the cell phone network data that may be of interest, including overall usage levels, spatial distribution of users, and the underlying social network.
- The *Simulation and Prediction System* (SPS) receives anomaly alerts from the DAS, produces hypotheses that describe the anomaly, and uses simulations in conjunction with streaming activity data to validate or reject the hypotheses. We also use the simulations resident in the SPS to generate our synthetic datasets described earlier.
- The *Historical Data Source* (HIS) is a repository of cellular network data that resides in secondary storage. This data is used to determine the base-line behavior of the network against which anomalies are detected and to periodically calibrate and update the DAS.

- The *Real-Time Data Source* (RTDS) is a real-time system that will receive transaction data directly from a cellular service provider. The RTDS is responsible for handling requests for streaming data from the DAS, SPS, and DDS and streaming incoming data to these components in a timely manner.

Figure 1 shows an architectural overview of the WIPER system. The RTDS and HIS will provide the bridge from the service provider and the WIPER system. The figure shows the flow of streaming data from the service provider through the RTDS, possibly by way of the HIS for development and training, and to the remaining components. Requests for streaming data from the RTDS occur via SOAP messages. SOAP messages are also used by the Detection and Alert System to inform the Simulation and Prediction system of potential anomalies in the streaming data.

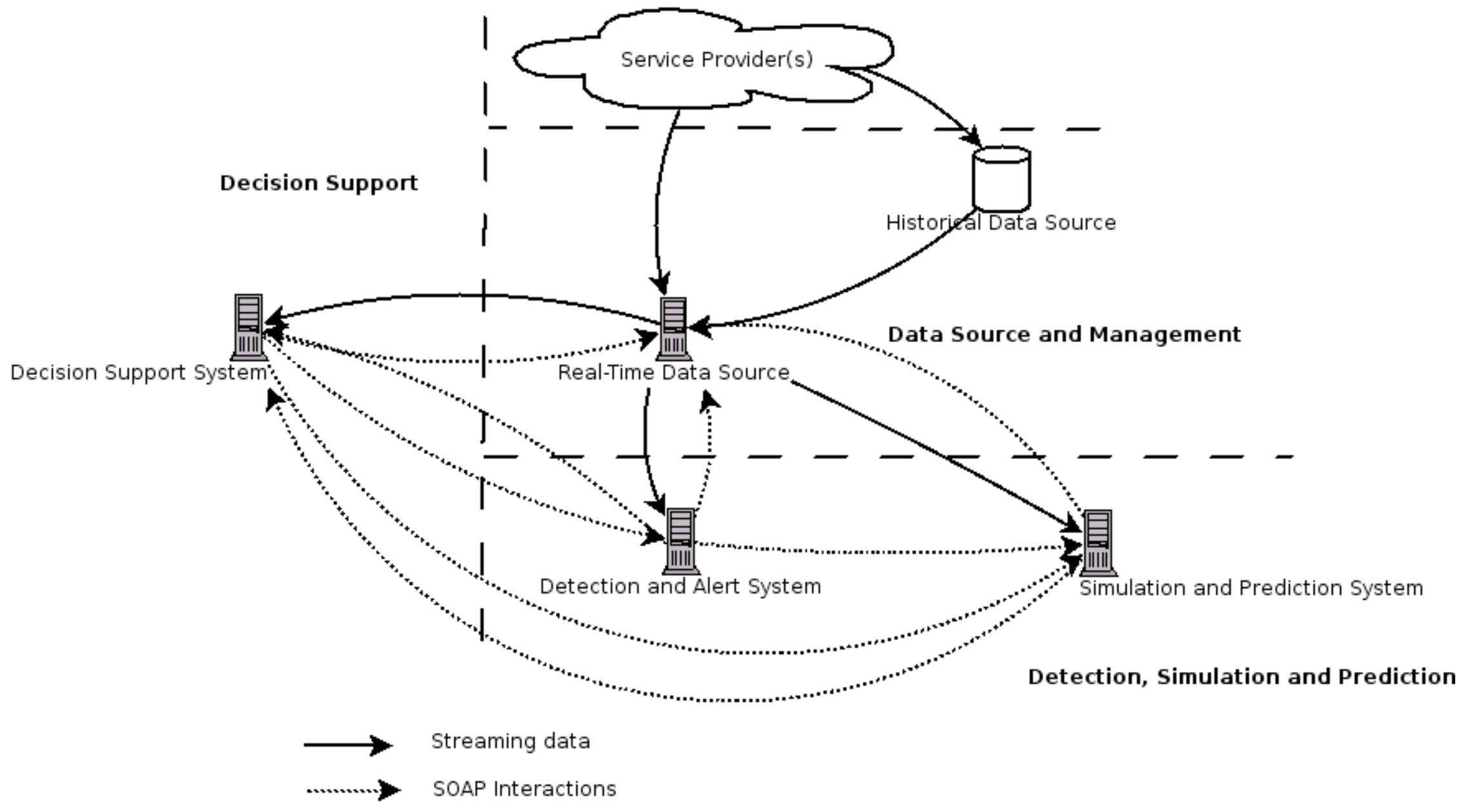

*Figure 1: WIPER system architecture.*

# The Detection and Alert System

The detection and alert system is designed to examine the streaming data from the cellular service provider for anomalous activity on two axes: call activity (the number of calls made in a fixed time interval), and spatial distribution (location and movement) of the cell phones based on calls made using them. Three data mining techniques have been implemented and evaluated for use in the Detection and Alert System of WIPER: 1) a model that uses a Markov modulated Poisson process technique, 2) a method for spatial analysis based on percolation theory, and 3) a method for spatial analysis using online hybrid clustering. These techniques and their adaptation to data mining of cell phone data for anomalies within the WIPER system are described below.

## *Call Activity Analysis using Markov Modulated Poisson Processes*

The most basic indicator of anomalous behavior in a cell phone network is an increase or a decrease in cell phone call activity within a given geographical area. This type of anomaly can be detected by monitoring a time series consisting of the number of calls made in disjoint time intervals of a fixed size, *e.g.* the number of calls made every 10 minutes. The Markov modulated Poisson process, which uses a Poisson process in conjunction with a hidden Markov model to identify anomalies in the data, is described by Ihler, Hutchins, and Smyth (2006, 2007) and is summarized below.

A Poisson process, which models the number of random events that occur during a sequence of time intervals, can be used to model the baseline behavior of such a time series: the number of events per time interval follows a Poisson distribution with an expected value of $\lambda$, the rate parameter for the process. In this model, the probability of $N$ events occurring in a time step is:

$$P(N;\lambda) = \frac{e^{-\lambda}\lambda^N}{N!} \qquad (5)$$

for $N$=0,1,… (Mitzenmacher & Upfal, 2005).

The standard Poisson process is not sufficient for modeling many real-world phenomena since the rate of many natural processes varies over time. In the case of human activities, there are often daily and weekly cycles, so the rate becomes a function of the day of the week and time of day. The overall average, , is the average rate over all time intervals and establishes the baseline rate of the process. The day effect, $\delta_{d(t)}$, $d(t) \in \{1 \ldots 7\}$, is the average rate over all time intervals for each day of the week, normalized such that the average day effect is 1, *i.e.* $\sum \delta_{d(t)} = 7$. The day effect expresses the call activity of the day relative to the overall average. The time of day effect, $\eta_{d(t),h(t)}$, $h(t) \in \{1 \ldots D\}$ is the average rate for each time interval for each day of the week. The time of day effect for each of the 7 days of the week is normalized such that average time of day effect for each day is 1, *i.e.* $\forall d(t), \sum \eta_{d(t),h(t)} = D$. The time of day effect expresses the call activity of the interval relative to the product of the overall average and the day effect.

The rate function for a Markov modulated Poisson process is

$$\lambda(t) = \lambda_0 \delta_{d(t)} \eta_{d(t),h(t)} \qquad (6)$$

To illustrate the components of the rate function, we compute the overall average rate, the day effect, and the time of day effect from two weeks of real cell phone data. Figure 2 shows each component of the rate function.

The Poisson process described above is used in conjunction with a hidden Markov model to identify anomalies in the call data. The hidden Markov model has two states for describing the current call activity: normal and anomalous. The transition probability matrix for the hidden

Markov model is

$$M = \begin{bmatrix} 1-A_0 & A_1 \\ A_0 & 1-A_1 \end{bmatrix} \qquad (7)$$

where each entry $m_{ij} \in M$ corresponds to the probability of moving from state $i$ to state $j$. Intuitively, $1/A_0$ is the expected time between anomalies and $1/A_1$ is the expected duration of an anomaly. Initially, we guess that $A_0 = 0.01$ and $A_1 = 0.25$. These guesses are updated based on the state sequence generated in each iteration.

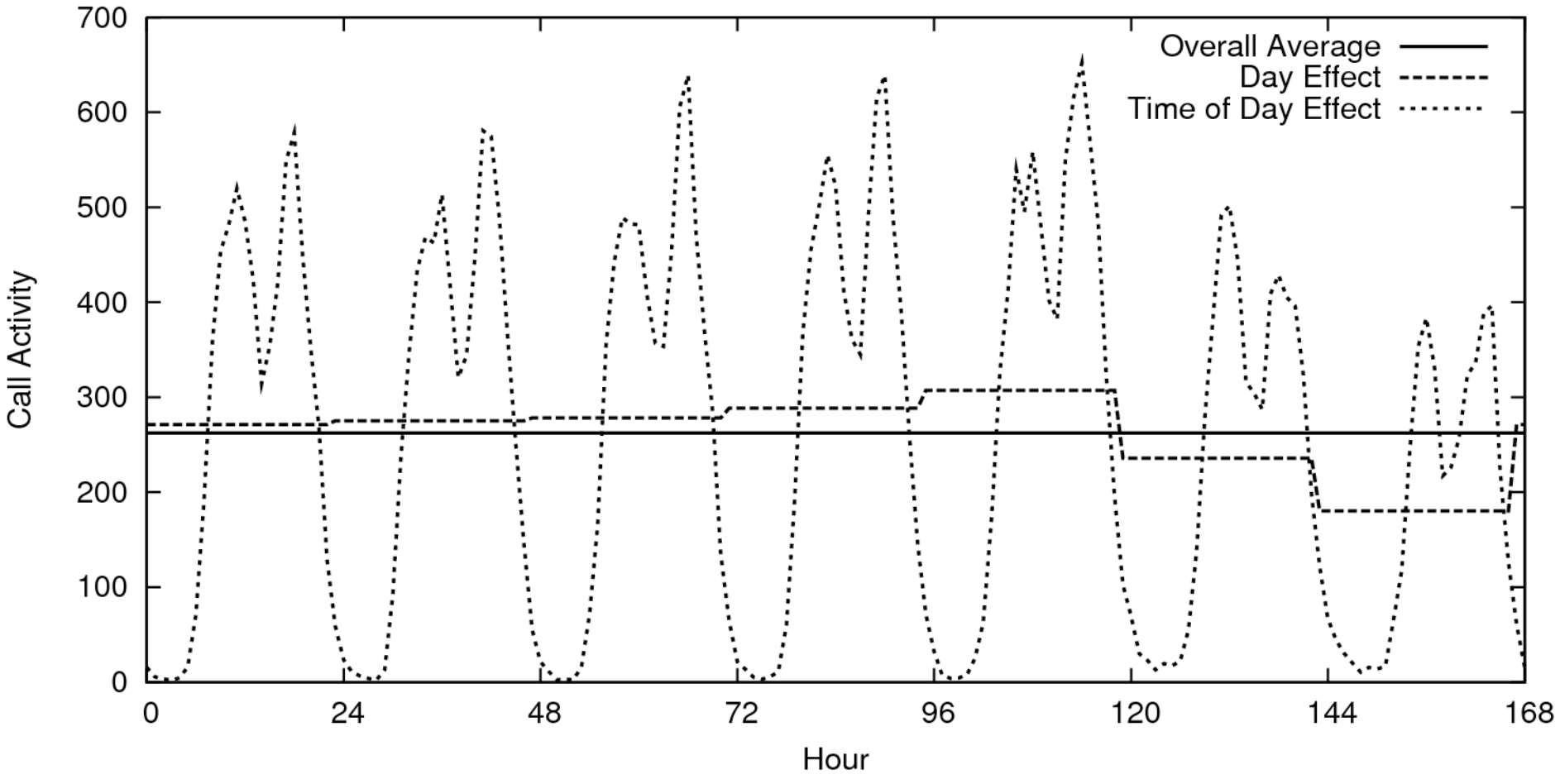

*Figure 2: The overall average rate ( $\lambda_0$), day effect combined with the overall average ( $\lambda_0\delta_{d(t)}$), and time of day effect combined with the overall average and the day effect ( $\lambda_0\delta_{d(t)}\eta_{d(t),h(t)}$).*

Anomalies are identified using the Markov Chain Monte Carlo method. For each iteration of the method, the forward-backward algorithm is used to determine a sample state sequence of the hidden Markov model. For each interval in the forward recursion $t:1,2,\ldots T$, the probability of each hidden Markov state is computed by

$$p(A(t) \mid N(t)) = \pi_0 \sum M \cdot p(A(t-1) \mid N(t-1)) p(N(t) \mid A(t)) \qquad (8)$$

where $\pi_0$ is the initial distribution of the Markov chain. If the hidden Markov model is in the normal state, the likelihood function, $p(N(t) \mid A(t))$, is simply the probability $N(t)$ is generated by the Poisson process at time $t$. If the hidden Markov model is in the anomalous state, the likelihood function takes into account the range of possible number of observations, $i \in \{0,1,\ldots N(t)\}$, beyond the expected number. The probability that $i$ of the $N(t)$ observations are normal is computed using a negative binomial distribution. Let $\mathrm{NBIN}(N,n,p)$ be the probability of $N$ observations given a negative binomial distribution with parameters $n,p$, and let this negative binomial distribution model the number of anomalous observations, $N(t)-i$, in an interval. The likelihood function is

$$p(N(t) \mid A(t)) = \begin{cases} P(N(t);\lambda(t)) & A(t)=0 \\ \sum_{i=0}^{N(t)} P(i,\lambda(t))\mathrm{NBIN}(N(t)-i;a^E,\frac{b^E}{1-b^E}) & A(t)=1 \end{cases} \tag{9}$$

where $a^E = 5$ and $b^E = 1/3$ are empirically determined parameters of the negative binomial distribution.

For each interval in the backward recursion, $t: T, T-1, \ldots 1$, samples are drawn from the conditional distribution $M' \cdot p(A(t) \mid N(t+1))$, where $M'$ is the inverse of the transition probability matrix, to refine the probability of the current state $t$.

Once the forward-backward algorithm has generated a sample hidden state sequence, the values of the transition probability matrix are updated using the empirical transition probabilities from the sample state sequence, and the process is repeated.

We apply this approach to two weeks of call activity data taken from our primary data set (i.e., actual call data), using 50 iterations of the Markov Chain Monte Carlo simulations described above to determine the probability of anomalous behavior for each 10 minute interval. Figure 3 shows the actual call activity and the call activity modeled by the Markov modulated Poisson process for two weeks of for a small town with 4 cell towers. Visual inspection of the graph indicates that the Markov modulated Poisson process models the real call activity well. We do not have information about any emergency events that may be present in this dataset; therefore, this figure shows the posterior probability of an anomaly at each time step in the lower frame based on the hidden Markov model. Note that on the last day of observation, the Markov modulated Poisson process identifies an anomaly corresponding to an observed call activity that is significantly higher than expected. Additionally, an anomaly is detected on the second Tuesday; however, we cannot see a major deviation from the expected call activity raising the possibility that this is a false positive. For each remaining interval, the posterior probability of an anomaly is no greater than 0.5. This analysis indicates that outliers in the call activity time series data can be identified using a Markov modulated Poisson process and could be useful as an alerting method to indicate possible anomalies and emergency events. Such a system would need a second stage of analysis to determine if the outlier is a true positive for an emergency event. These detected anomalies trigger an alert that is sent to the Decision Support System and the Simulation and Prediction System of the WIPER system. Yan, Schoenharl, Pawling, and Madey (2007) describe in greater detail this application of a Markov modulated Poisson process to the problem of detecting outliers and anomalies in call activity data.

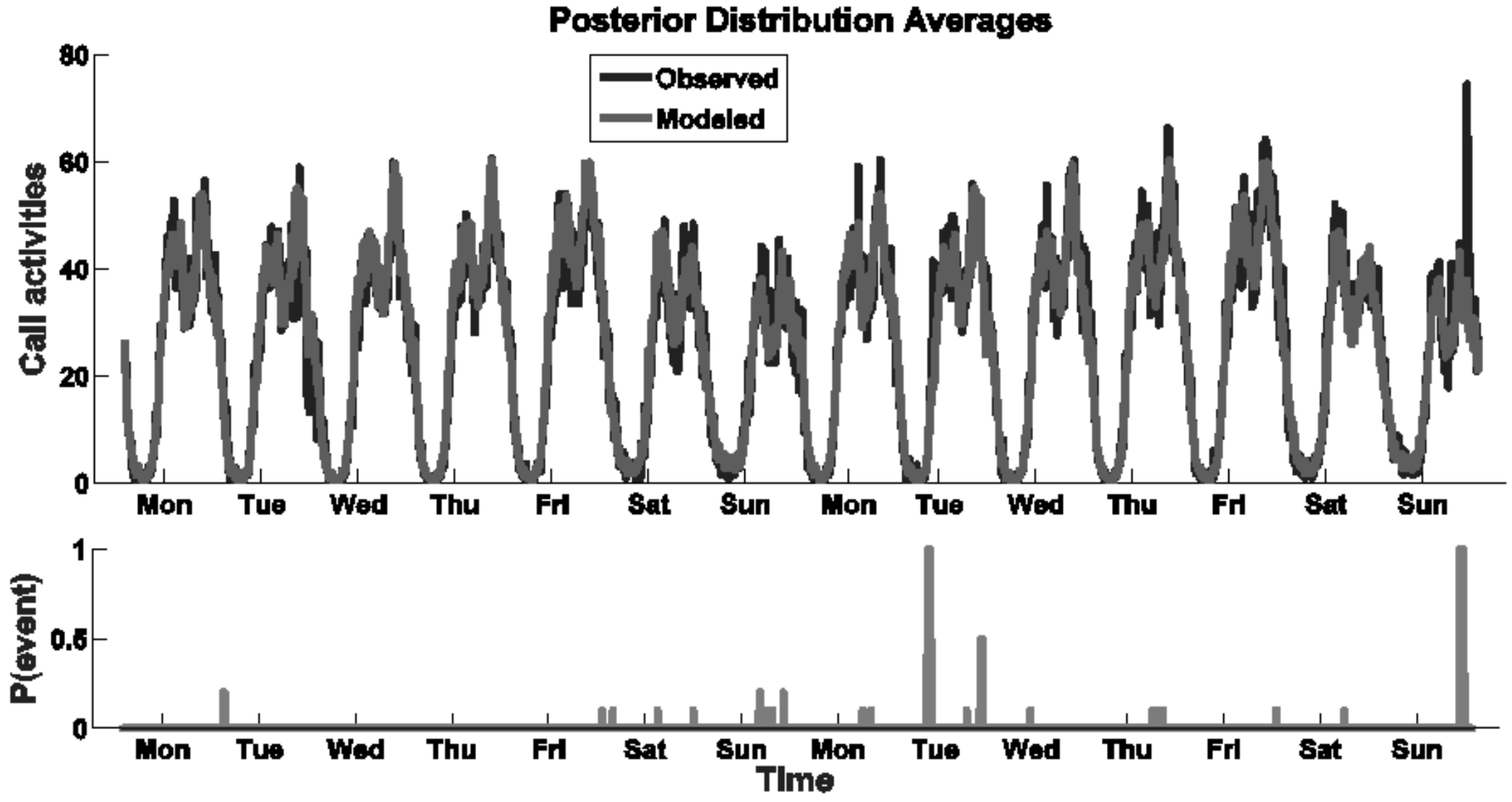

*Figure 3: This figure shows the result of using a Markov modulated Poisson process to detect anomalies in 2 weeks of call activity. The top frame shows the expected and observed number of calls for each time interval, and the bottom frame shows the probability that the observed behavior is anomalous at each time step.*

## *Spatial Analysis using Percolation Theory*

We have determined that models based on percolation theory can be used to detect spatial anomalies in the cell phone data. The geographical area covered by the data set is divided into a two dimensional lattice, and the call activities through the towers within each cell of the lattice are aggregated. The normal activity for each cell is defined by the mean and standard deviation of the call activity, and a cell is in an anomalous state when its current observed call activity deviates from the mean by some factor, $l$, of the standard deviation. In the percolation theory model, neighboring anomalous sites are connected with an edge. When an anomaly occurs in the cell phone network, the number of clusters and the distribution of cluster sizes are statistically different from those that arise due to a random configuration of connected neighbors. In contrast, when the cell phone network is behaving normally, the number of clusters and distribution of cluster sizes match what is expected. Candia et al. (2008) provide a more detailed discussion of percolation theory and how the spatial anomalies of the cell phone data can be detected.

### *Spatial Analysis using Online Hybrid Clustering*

We have evaluated a hybrid clustering algorithm for online anomaly detection for the WIPER system. This hybrid algorithm is motivated by the fact that streaming algorithms for clustering, such as those described by Guha et al (2003) and Aggarwal et al (2003), require *a priori* knowledge of the number of clusters. Due to the dynamic nature of the data stream, we believe that an algorithm that dynamically creates new clusters as needed, such as the leader algorithm, is more appropriate for this application. However, we also believe that the leader algorithm is too inflexible since it produces clusters of a constant size.

The hybrid algorithm combines a variant of the leader algorithm with $k$-means clustering to overcome these issues. The basic idea behind the algorithm is to use $k$-means to establish a set of clusters and the leader algorithm in conjunction with statistical process control to update the clusters as new data arrives. For detecting anomalies in the spatial distribution of call activity, the feature vectors consist of the call activities for each cell tower in the area of interest.

Statistical process control aims to distinguish between "assignable" and "random" variation. Assignable variations are assumed to have low probability and indicate some anomaly in the underlying process. Random variations, in contrast, are assumed to be quite common and to have little effect on the measurable qualities of the process. These two types of variation may be distinguished based on the difference in some measure on the process output from the mean, $\mu$, of that measure. The threshold is typically some multiple, $l$, of the standard deviation, $\sigma$. Therefore, if the measured output falls in the range $\mu \pm l\sigma$, the variance is considered random; otherwise, it is assignable (Bicking & Gryna, Jr., 1979).

The algorithm represents the data using two structures: the cluster set and the outlier set.

To save space, the cluster set does not store the examples that make up each cluster. Instead, we use the summarization approached described by Zhang, Ramakrishnan & Livny (1996), where each cluster is summarized by the sum and sum squared values of its feature vectors along with the number of items in the cluster. The outlier set consists of the examples that do not belong to any cluster. The means and the standard deviations describe the location and size of the clusters, so clusters are only accepted when they contain some minimum number of examples, $m$, such that these values are meaningful. The algorithm periodically clusters the examples in the outlier set using $k$-means. Clusters that contain at least $m$ items are reduced to the summary described above and added to the cluster set. If a new data point is within the threshold, $l\sigma$, of the closest cluster center, it is added to the cluster and the summary values are updated. Otherwise, it is placed in the outlier set.

By using mean values as the components of the cluster center and updating the centers whenever a new example is added to a cluster, the algorithm can handle a certain amount of concept drift. At the same time, the use of statistical process control to filter out anomalous data prevents the cluster centers from being affected by outlying points. This algorithm does not require *a priori* knowledge of the number of clusters, since new clusters will form as necessary.

This approach does have some drawbacks. There are cases in which the $k$-means clustering component will fail to produce any clusters of sufficient size; however, we have successfully used this algorithm on data vectors containing usage counts of 5 services provided by a cellular communication company at one minute intervals and simulated spatial data. This hybrid clustering algorithm used for online anomaly detection is described in more detail in Pawling, Chawla, and Madey (2007)

# Discussion

## *Results and Limitations*

WIPER is a proof-of-concept prototype that illustrates the feasibility of dynamic data driven application systems. It has been shown that anomalies in real world data can be detected using Markov modulated Poisson processes (Yan et al, 2007) and percolation theory (Candia et al, 2008). The hybrid clustering algorithm has been evaluated using synthetic spatial data generated from simulations based on real-world data with promising results.

The detection and alert system assumes that emergency events are accompanied by a change in underlying call activity. In cases where this does not hold, the system will fail to identify the emergency. Additionally, in cases where the underlying call activity changes very gradually, the system may fail to detect the situation.

In its current state, WIPER can only identify that an anomaly has occurred, it cannot make any determination of its cause. Therefore, the system cannot distinguish between elevated call activity due to an emergency, such as a fire, from a benign event such as a football game. The WIPER system is a laboratory prototype with no immediate plans for deployment. Laboratory tests have demostrated that the individual components perform as desired and that the multiple modules can work in a distributed manner using SOAP messaging.

## *Data Mining and Privacy*

As the fields of database systems and data mining advance, concerns arise regarding their effects on privacy. Moor (1997) discusses a theory of privacy in the context of "greased" data, data that is easily moved, shared, and accessed due to advances in electronic storage and information retrieval. Moor argues that as societies become large and highly interactive, privacy becomes necessary for security.

"Greased" data is difficult to anonymize because it can be linked with other databases, and there have been cases where data has been "de-identified" but not "anonymized". That is, all identifying fields, such as name and phone number, have been removed or replaced but at least one person's identity can be determined by linking the records to other databases. In these cases, the remaining fields uniquely identify one or more individuals (Sweeney, 1997). With the development of new technologies, data sets thought to be anonymized when collected can become de-anonymized as additional data sets become available in the future. Thus anonymizing "greased" data is extremely difficult. (National Research Council, 2007).

Geographic Information Systems (GIS) provide additional data against which records can be linked. For safety reasons, some governments require that telecommunication companies be able to locate cell phones with some specified accuracy so that people calling for emergency services can be quickly located. Emergency responders can easily find a phone by plotting the location on maps using GIS technology. This method of locating phones can also be used to provide subscribers with location-based services, or it can be used to track an individual's movements (Armstrong, 2002).

A significant issue that arises in the discussion of data mining and privacy is the difficulty of precisely defining privacy. Solove (2002) surveys the ways in which privacy has been conceptualized throughout the history of the U.S. legal system, and points out serious shortcomings of each. Complicating the issue further is the fact that ideas of privacy are determined by culture and are constantly evolving, driven in part by advances in technology (Armstrong & Ruggles, 2005).

Clifton, Kantarcioglu, and Vaidya (2002) describe a framework of privacy for data mining. This paper looks at two types of privacy: individual privacy, which governs information about specific people, and corporate privacy, which governs information about groups of people. In general, individual privacy is maintained, from a legal standpoint, if information cannot be tied to a single individual. Corporate privacy aims to protect a data set, which includes protecting the results of analysis of the data. In a follow-up paper, Kantarcioglu, Jin, and Clifton (2004) propose framework for measuring the privacy preserving properties of data mining results. This framework assumes that the data includes fields that are public, sensitive, and unknown but not sensitive. The framework provides measures of how well the sensitive fields are protected against various attacks using the classifier, such as attempting to infer the values of sensitive fields using public fields.

In response to privacy concerns relating to data mining, researchers are developing data mining methods that preserve privacy. Agrawal and Srikant (2000) propose an approach to classification that achieves privacy by modifying values such that a reliable model may be built without knowing the true data values for an individual. Two methods are used for modifying attribute values: (1) value-class membership is essentially a discretization method that aggregates values into intervals, each of which has a single associated class, and (2) value distortion in which random noise is added to the real value. In the case of value distortion, the data distribution is recovered based on the result of the distortion and the distribution of the distorting values, but the actual attribute values remain hidden.

Lindell and Pinkas (2002) describe a privacy preserving data mining protocol that allows two parties with confidential databases to build a data mining model on the union of the databases without revealing any information. This approach utilizes homomorphic encryption functions. Homomorphic encryption functions allow computations on encrypted values without revealing the actual values. Benaloh (1994) and Paillier (1999) describe additively homomorphic public key encryption functions. Let $E$ be an encryption function and $x$ and $y$ be plaintext messages. If $E$ is additively homomorphic, $E(x)$ and $E(y)$ can be used to compute $E(x+y)$ without revealing $x$ or $y$. This classification method assumes "semi-honest" parties that correctly follow the protocol but try to obtain further information from the messages passed during the computation.

Friedman, Schuster, and Wolff (2006) describe a decision tree algorithm that produces $k$-anonymous results with the goal of preventing linking attacks that use public information and a classifier to infer private information about an individual. They describe a method for inducing a decision tree in which any result from the decision tree can be linked to no fewer than $k$ individuals.

The nature of the phone data set raises some concerns about privacy issues in relation to our work. Data stored by service providers allows fairly detailed tracking of individuals based on the triangulation of radio signals received by cell towers from phones, as well as the capability to identify an individual's current location. A major concern is the potential for abuse of this technology by the government and law enforcement, especially considering that there is no consensus on what level of evidence is required to gain this information from cellular service providers. Some judges require law enforcement to show probable cause before allowing this data to be accessed, while others view this information as public, since cell phone users choose to keep their device powered on (Nakashima, 2007).

Compounding this concern is the fact that following the terrorist attacks on September 11, 2001 in the U.S., a number of U.S. airlines provided the U.S. government with their passenger records, in direct violation of their own privacy policies. The courts did not accept arguments that this was a breach of contract since no evidence was provided that this breach of contract caused any harm. Solove (2007) argues that the harm here is a loss of trust in companies and the rise of an imbalance in power, since, apparently once a company has information about an individual, the individual loses control over that information completely. In a similar, and more widely known case, U.S. telecommunication companies provided the U.S. government with call records for their subscribers, violating a long held tradition of only releasing customer information when ordered to do so by a court (Cauley, 2006).

In the European Union, privacy is viewed as a Human Right. As a result, the privacy laws are much more comprehensive and are extensive in their coverage of both private and public institutions. In 1968, the Council of Europe discussed the impact of scientific and technological advances on personal privacy, with a focus on bugging devices and large-scale computerized storage of personal information. This discussion led to an evaluation of the adequacy of privacy protection provided by the national laws of member states given recent advances in technology, and preliminary reports indicated that improvement was needed. In 1973 Sweden passed the Data Protection Act requiring governmental approval and oversight of any “personal data register”. This was followed by similar legislation in Germany, France, Denmark, Norway, Austria, and Luxembourg by 1979 (Evans, 1981) and the European Data Privacy Directive in 1995 (European Parliament and Council of the European Union, 1995).

The European Data Privacy Directive requires “adequate” data privacy protections be in place before personal data of European Union citizens can be exported to a country outside the Union (European Parliament and Council of the European Union, 1995). In general, the United States does not provide an “adequate” level of protection; however, the U.S. Department of Commerce developed the “Safe Harbor” program that allows American businesses to continue receiving data from Europe by certifying that their data protection policies meet the requirements of the European Union (Murray, 2001).

“Safe Harbor” requires that companies notify customers of how their personal data is used, provide customers with ways in which to make inquiries and lodge complaints relating to their personal information held by the company, and provide customers with information about data sharing policies along with avenues for allowing the customer to limit the use and sharing of their personal data. In cases where personal data is shared with third parties or used for a new purpose, users must be given an opportunity to “opt out”, and in cases where this data is particularly sensitive, *e.g.* medical or health data, religious affiliation, or political views, the customer must “opt in” before the data can be shared (Murray, 2001).

Issues of data security, integrity, and access are also addressed by “Safe Harbor”. Companies in possession of personal data are required to take “reasonable precautions” to prevent security compromises, including unauthorized access, disclosure, and alteration of the data. Data integrity refers to the relevance and reliability of the data. Companies must have a specific use for each item of personal information in order to obtain it and may not use that data for any other purpose without consent of the individual described by the data. Finally, users are required to have access to their personal data possessed by the company and the company must provide mechanisms that allow individuals to correct any inaccuracies in the data or request its deletion (Murray, 2001).

## *Future Directions*

Several tasks remain to be completed on this project: incorporation of link mining and social network analysis into the stream mining component of the WIPER system, the development of a better understanding of the relationship between outliers, anomalies, and emergencies in our data, and finally the field testing of the system, both with emergency managers within an emergency operations center and with a live stream from a cellular carrier.

Much of the previous work in identifying anomalies in graphs is based on subgraph matching; however, these approaches tend to be computationally expensive. Another possibility is clustering graphs based on some vector of metrics. Like the call activity, graph properties such as assortativity and clustering coefficient exhibit daily and weekly periodic behavior. It may be possible to identify outliers and classify emergency situations using vectors of graph metrics computed on graphs built from a sliding window of call transactions.

There are still important issues that must be resolved. It is not clear what graph properties should be used, and the appropriate window size must be determined. Unsupervised feature selection methods (Dy & Brodley, 2004, Mitra, Murthy, & Pal, 2002) from data mining may be used to identify the best set of graph properties from those that can be computed quickly.

## Summary

In this chapter, we have described the detection and alert component of the Wireless Phone-based Emergency Response System, a proof of concept dynamic data-driven application system. This system draws from research in data mining and percolation theory to analyze data from a cell phone network on multiple axes of analysis to support dynamic data-driven simulations.

## Acknowledgment

This material is based upon work supported in part by the National Science Foundation,

DDDAS program, under Grant No. CNS-0540348, ITR program (DMR-0426737), and IIS-0513650 program, the James S. McDonnell Foundation 21st Century Initiative in Studying Complex Systems, the U.S. Office of Naval Research Award N00014-07-C, the NAP Project sponsored by the National Office for Research and Technology (CKCHA005). Data analysis was performed on the Notre Dame Biocomplexity Cluster supported in part by NSF MRI Grant No DBI-0420980.